\documentclass{sf2a-conf2014}
\usepackage{graphicx}
\usepackage{hyperref}
\usepackage[]{natbib}  
\usepackage{epstopdf}
\usepackage{bm}
\usepackage{amsmath}
\usepackage{amssymb}

\def\BibTeX{{\rm B\kern-.05em{\sc i\kern-.025em b}\kern-.08em
    T\kern-.1667em\lower.7ex\hbox{E}\kern-.125emX}}
\bibpunct{(}{)}{;}{a}{}{,}  

\begin{document}

\TitreGlobal{SF2A 2014}


\title{Understanding tidal dissipation in gaseous giant planets : the respective contributions of their core and envelope}

\runningtitle{Understanding tidal dissipation in gaseous giant planets}

\author{M. Guenel}\address{Laboratoire AIM Paris-Saclay, CEA/DSM/IRFU/SAp - Universit\'e Paris Diderot - CNRS, 91191 Gif-sur-Yvette, France}
\author{S. Mathis$^1$}
\author{F. Remus}\address{IMCCE, Observatoire de Paris, CNRS UMR 8028, UPMC, USTL, 77 Avenue Denfert-Rochereau, 75014 Paris, France}



\setcounter{page}{237}


\maketitle


\begin{abstract}
Tidal dissipation in planetary interiors is one of the key physical mechanisms that drive the evolution of star-planet and planet-moon systems. New constraints are now obtained both in the Solar and exoplanetary systems. Tidal dissipation in planets is intrinsically related to their internal structure. In particular, fluid and solid layers behave differently under tidal forcing. Therefore, their respective dissipation reservoirs have to be compared. In this work, we compute separately the contributions of the potential dense rocky/icy core and of the convective fluid envelope of gaseous giant planets, as a function of core size and mass. We then compare the associated dissipation reservoirs, by evaluating the frequency-average of the imaginary part of the Love numbers $k^2_2$ in each region. We demonstrate that in general both mechanisms must be taken into account.
\end{abstract}



\section{Introduction}

The orbital and rotational evolution of a close-in planet around its host star or of a moon around a planet is strongly dependent on the tidal dissipation inside each body \citep[][]{GS1966}. However, the response of fluid and solid planetary layers to tidal excitation is not well-understood yet, as well as the associated dissipative processes which are very different in each type of region \citep[e.g.][]{MathisRemus2013,ADLPM2014}. For these reasons, there is a strong need for reliable calculations of the dissipation rate of the energy of tidal displacements in each kind of planetary layer.

Recent progress on observational constraints was obtained using high-precision astrometry measurements in the solar system \citep{Laineyetal2009,Laineyetal2012} especially for Jupiter and Saturn, and space-based high-resolution photometry for exoplanetary systems \citep{Albrechtetal2012}. These results showed that there may be a strong tidal dissipation in gaseous giant planets, and its smooth dependence on the tidal frequency in the case of Saturn indicates that the inelastic dissipation in their central dense core may be strong \citep[e.g.][]{RMZL2012,Remus2014,Storchetal2014}. However, on one hand, the mass, the size, and the rheology of these cores are still unknown. On the other hand, inertial waves, whose restoring force is the Coriolis acceleration, may be excited by tides in the surrounding fluid convective envelope. Moreover, it seems that turbulent friction acting on these waves can be strong too \citep[e.g.][]{OgilvieLin2004,Ogilvie2013}. As a consequence, it is necessary to develop new models that take into account the appropriate dissipative mechanisms, so that we can predict how much energy each type of layer can dissipate. This should be achieved not only for gaseous giant planets but for all multi-layer planets, that may consist of differentiated solid and fluid layers.

In this work, we used a simplified two-layer model that accounts for the internal structure of gaseous giant planets. We used the frequency-dependent Love number to evaluate the reservoirs of dissipation in both regions, in a way introduced by \cite{Ogilvie2013}. It allows us to give the first direct comparison of the respective strengths of different dissipative mechanisms occurring in a given planet. In sec. \ref{sec:modelling}, we describe the main characteristics of our simplified planetary model. Next, we recall the method we used to compute the reservoirs of dissipation that is a result of viscoelastic dissipation in the core \citep{RMZL2012,Remus2014} and of turbulent dissipation in the fluid envelope \citep{Ogilvie2013}. In sec. \ref{sec:comparison}, we explore their respective strength for possible values for the parameters of our two-layer model. Finally, we discuss our results and the potential applications of this method.

\section{Modelling tidal dissipation in gaseous giant planets}

\label{sec:modelling}
\subsection{The two-layer model}

\begin{figure}[ht!]
\centering
\includegraphics[width=0.5\textwidth]{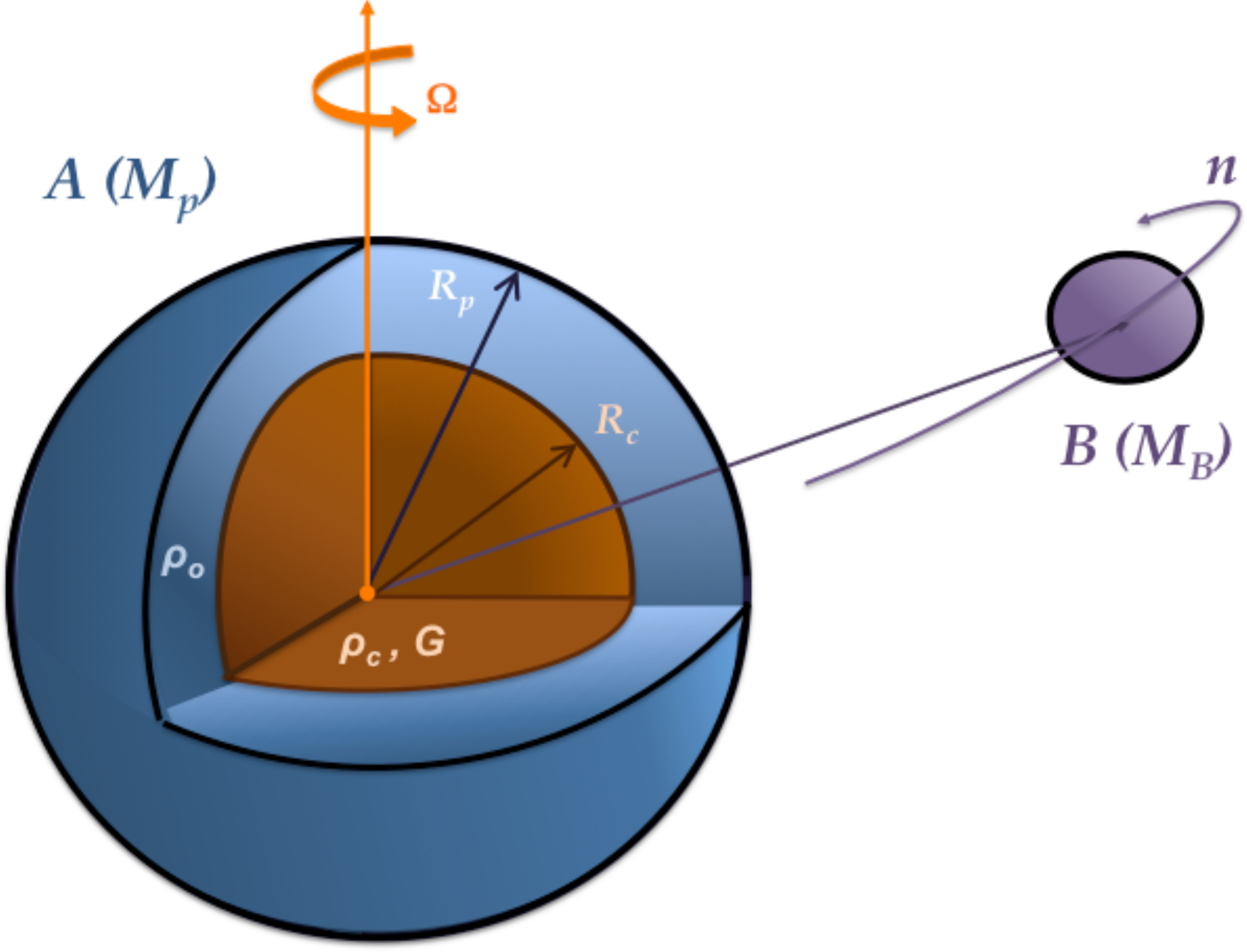}
\caption{The two-layer model.}
\label{GMR_fig1}
\end{figure}

This model features a central planet A of mass $M_p$ and mean radius $R_p$ assumed to be in moderate solid-body rotation $\Omega$ with $\epsilon^2 \equiv \Omega^2/ \sqrt{\mathcal{G} M_p / R_p^3} \ll 1$ (see fig. \ref{GMR_fig1}). In this regime, the Coriolis acceleration, which scales as $\Omega$, is taken into account while the centrifugal acceleration, which scales as $\Omega^{2}$ is neglected. The planet A has a rocky (or icy) solid core of radius $R_c$, density $\rho_c$ and rigidity $G$ that is surrounded by a convective fluid envelope of density $\rho_o$. Both regions are assumed to be homogeneous for the sake of simplicity. Finally, a point-mass tidal perturber B of mass $M_B$ is orbiting around A with a mean motion $n$.

\subsection{Mechanisms of dissipation}

The time-dependent tidal potential exerted by the companion leads to two different dissipation mechanisms. In the following, we detail how they operate and the hypotheses we used to evaluate their respective strength.

\begin{itemize}
\item First, we consider the viscoelastic dissipation in the solid core, for which we assume that the rheology follows the linear rheological model of Maxwell with a rigidity $G$ and a viscosity $\eta$ ; we also assume that the surrounding envelope is inviscid and only applies hydrostatic pressure and gravitational attraction on the core.

\begin{figure}[ht!]
\centering
\includegraphics[width=0.45\textwidth]{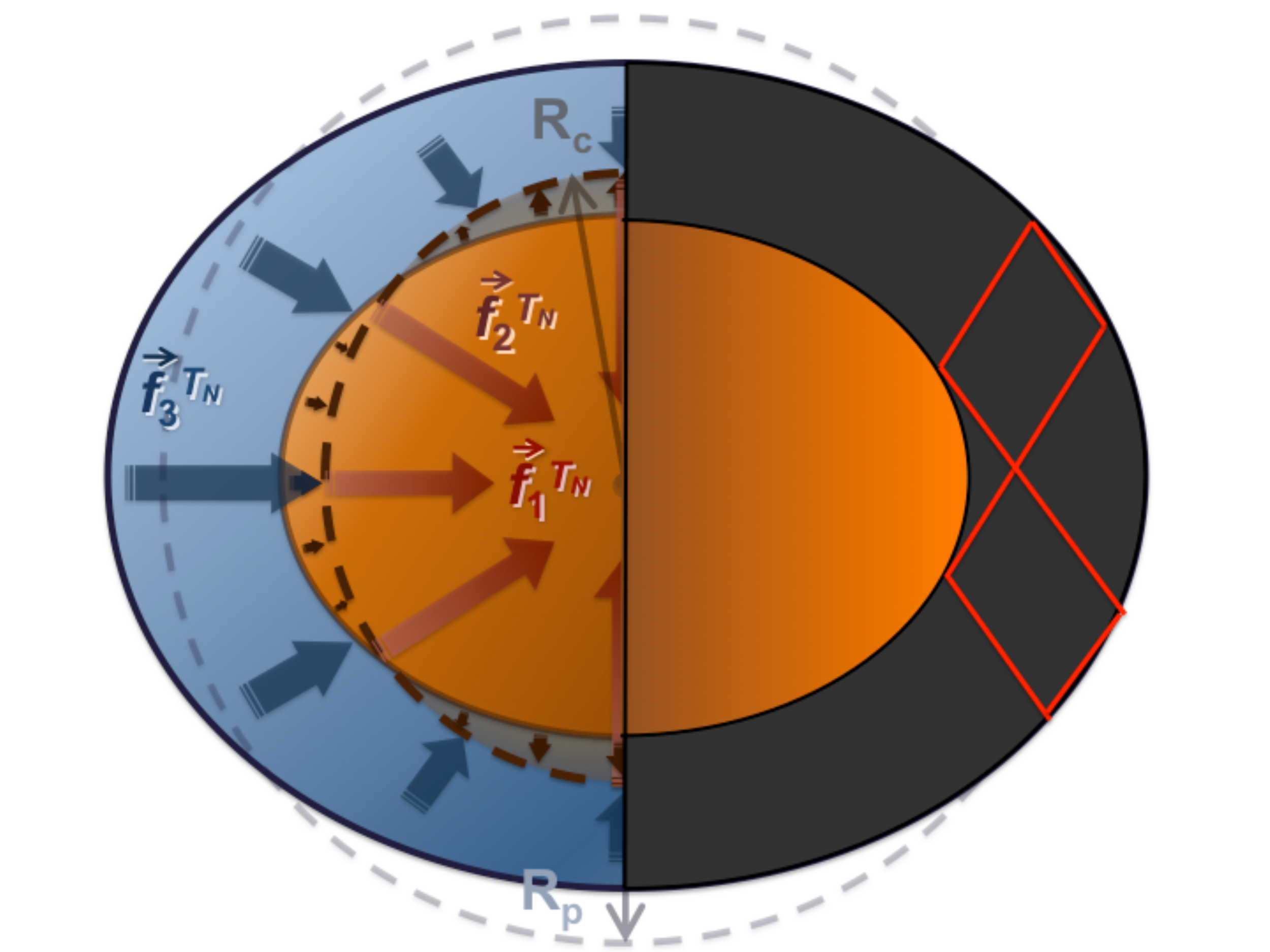}
\caption{{\bf Left~:} Gravitational forces ($\vec{f}_1$), internal constraints ($\vec{f}_2$) and hydrostatic pressure ($\vec{f}_3$) acting on the solid core, which is deformed by the tidal force exerted by the companion. {\bf Right~:} Attractor formed by a path of characteristics of inertial waves.}
\label{GMR_fig2}
\end{figure}

\item Then, the turbulent viscosity in the fluid convective envelope dissipates the kinetic energy of tidal inertial waves propagating in that region. The restoring force of inertial waves is the Coriolis acceleration and their frequency is smaller than the Coriolis frequency : $\omega \in [-2\Omega,2\Omega]$. Moreover, their kinetic energy may concentrate and form shear layers around attractor cycles, which leads to enhanced damping by turbulent viscosity. In order to compute it, the core is assumed to be perfectly rigid.
\end{itemize}

\subsection{Evaluation of the tidal dissipation reservoirs}

We compute for each of these mechanisms the "reservoir of dissipation", a weighted frequency-average of the imaginary part of the Love number $k^2_2(\omega) = {\Phi^2_2}'/U^2_2$ (which is the ratio between the $Y^2_2$-components of the Eulerian perturbation $\Phi'$ of the self-potential of body A, and of the tidal potential $U$) defined as :
\begin{equation}
\int_{-\infty}^{+\infty} \! {\rm Im} \left[k_2^2(\omega)\right] \,\frac{\mathrm{d}\omega}{\omega} = \int_{-\infty}^{+\infty} \! \frac{\left| k_2^2(\omega) \right|}{Q_{2}^{2}(\omega)} \,\frac{\mathrm{d}\omega}{\omega},
\label{eq:integralk22}
\end{equation}
where $Q^2_2(\omega)$ is the corresponding tidal quality factor.

\begin{itemize}
\item We find for the viscoelastic dissipation mechanism \citep[see][]{RMZL2012,Remus2014,GMR2014}:
\begin{equation}
\int^{+\infty}_{-\infty} \! {\rm Im} \left[k_2^2(\omega)\right] \,\frac{\mathrm{d}\omega}{\omega}  = \frac{\pi \,G \left(3 + 2\, \alpha\right)^2 \beta\, \gamma}{\delta \left(6 \, \delta+4\,\alpha\,\beta\,\gamma\, G\right)},
\end{equation}
where $\alpha,\beta$ and $\delta$ are positive functions of the aspect and density ratios $\left( R_c/R_p,\rho_o/\rho_c\right)$, whereas $\gamma$ only depends on $R_c$ and $\rho_c$. This result is remarkably independent of the viscosity $\eta$ while ${\rm Im}\left[k_2^2(\omega)\right]$ is not.

\item Meanwhile, \cite{Ogilvie2013} provides us for inertial waves :
\begin{equation}
\int^{+\infty}_{-\infty} \! {\rm Im} \left[k_2^2(\omega)\right] \,\frac{\mathrm{d}\omega}{\omega} = \frac{100 \pi}{63} \epsilon^2 \frac{\left( R_c/R_p \right)^5}{1-\left( R_c/R_p \right)^5} \times\left[ 1+ \frac{1-\rho_o / \rho_c}{\rho_o / \rho_c} \left( R_c/R_p \right)^3 \right] {\left[ 1+ \frac{5}{2} \frac{1-\rho_o / \rho_c}{\rho_o / \rho_c} \left( R_c/R_p \right)^3 \right]}^{-2}.
\label{eq:imk22ogilvie}
\end{equation}
\end{itemize}

\begin{figure}[ht!]
\centering
\includegraphics[width=0.56\textwidth]{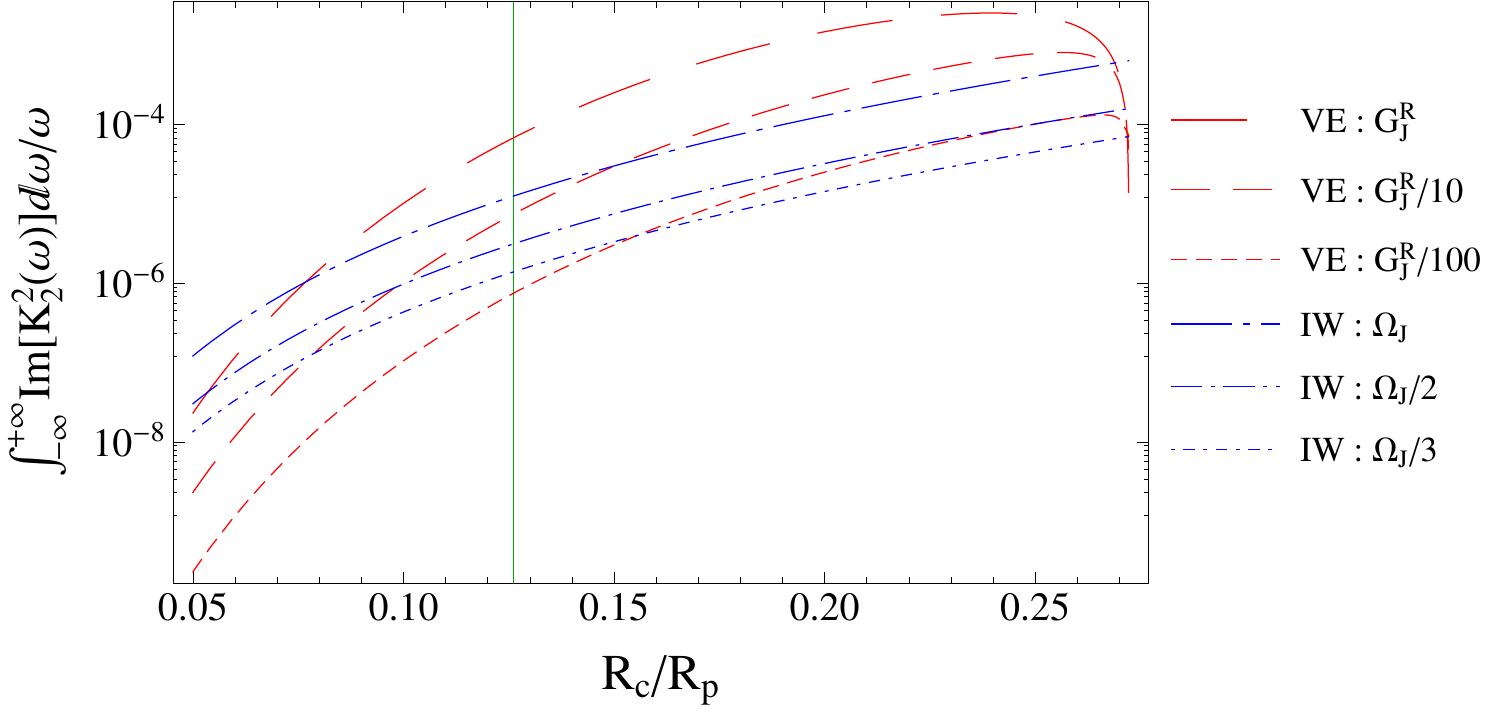}
\includegraphics[trim=25 0 100 0,clip,width=0.4\linewidth]{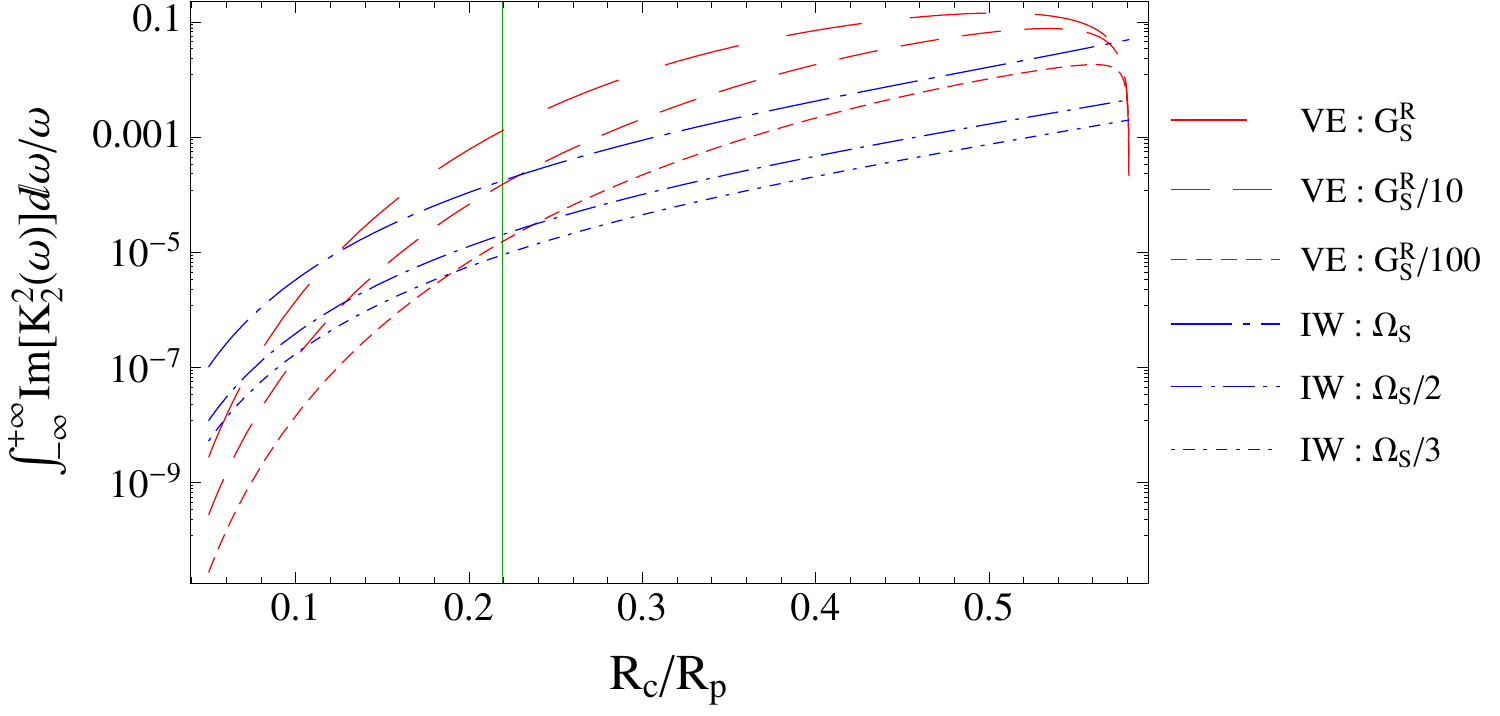}
\includegraphics[width=0.56\linewidth]{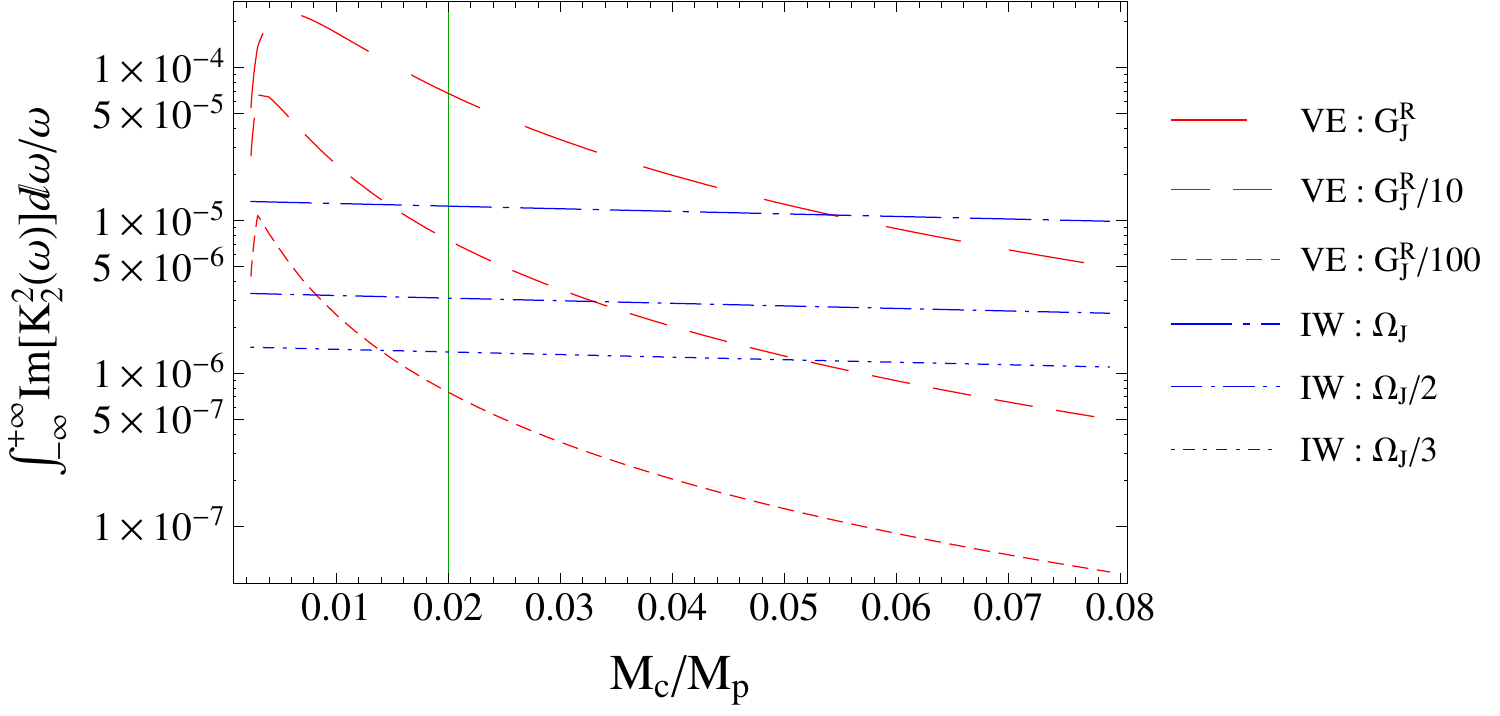}
\includegraphics[trim=25 0 100 0,clip,width=0.4\linewidth]{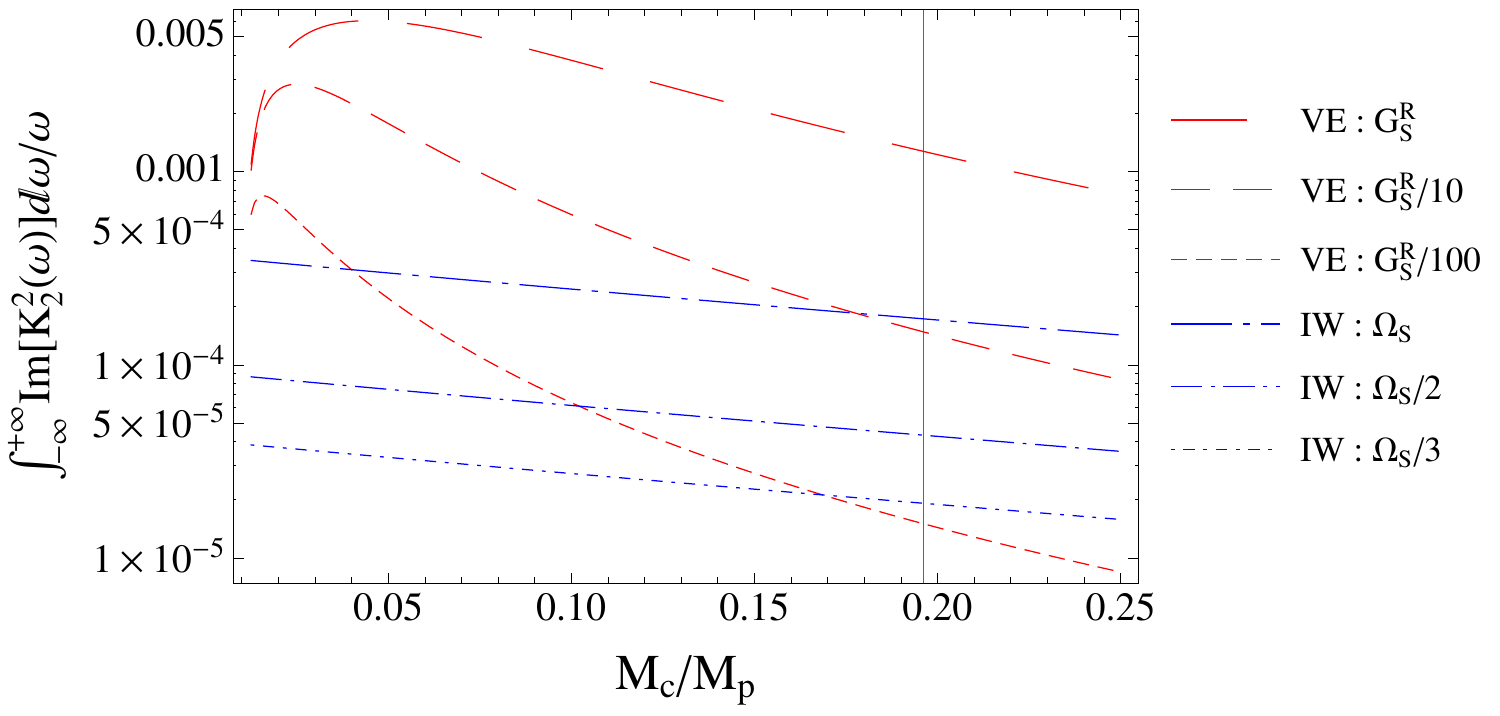}
\caption{{\bf Left~:} Dissipation reservoirs for the viscoelastic (VE) dissipation in the core (red curve) and the turbulent friction acting on inertial waves (IW) in the fluid envelope (blue curves) in Jupiter- (above) and Saturn-like planets (below) as a function of $R_c/R_p$, $\Omega$, and $G$, with fixed $R_p$ and $M_p$. We use the values $M_c/M_p = \left\{0.02,0.196\right\}$ for Jupiter and Saturn respectively. The vertical green line corresponds to $R_c/R_p = \left\{0.126,0.219\right\}$. {\bf Right~:} Similar to the left-side but now as a function of $M_c/M_p$ with fixed $M_p$ and $R_p$. We adopt $R_c/R_p = \left\{0.126,0.219\right\}$ for Jupiter and Saturn respectively. The wide $M_c$-ranges [1,3 - 25] $M_{\oplus}$ for Jupiter and [2 - 24] $M_{\oplus}$ for Saturn cover the values considered possible by various internal structure models \citep{Guillot1999,Hubbardetal2009}. The vertical green line corresponds to $M_c/M_p = \left\{0.02,0.196\right\}$.}
\label{GMR_fig34}
\end{figure}

\section{Comparison of the two dissipation mechanisms}
\label{sec:comparison}

\begin{itemize}
\item Our goal is to compare quantitatively the respective strength of the two dissipative mechanisms in order to determine if one of them can be neglected in gaseous giant planets similar to Jupiter and Saturn. Their respective mass and radius are $M_p=\left\{317.83,95.16\right\}M_{\oplus}$ and $R_p=\left\{10.97,9.14\right\}R_{\oplus}$ ($M_{\oplus}=5.97\,10^{24}$ kg and $R_{\oplus}=6.37\,10^3$ km being the Earth's mass and radius). Their rotation rate is $\Omega_{\left\{{\rm J,S}\right\}}=\left\{1.76\,10^{-4},1.63\,10^{-4}\right\}{\rm rad}\cdot{\rm s}^{-1}$. Internal structure models for these bodies are still not well constrained \citep{Guillot1999,Hubbardetal2009}. This is why we choose to explore wide ranges of core radii (left) and core masses (right) in fig. \ref{GMR_fig34}.

\item We choose to use as a reference $G_{\left\{{\rm J,S}\right\}}^{\rm R}=\left\{4.46\,10^{10},1.49\,10^{11}\right\}\,{\rm Pa}$ that allows the viscoelastic dissipation model to match the dissipation measured by \cite{Laineyetal2009, Laineyetal2012} in Jupiter at the tidal frequency of Io and in Saturn at the frequency of Enceladus (with $\eta_{\left\{{\rm J,S}\right\}}=\left\{1.45\,10^{14},5.57\,10^{14}\right\}\,{\rm Pa}\cdot{\rm s}$).

\item Figure \ref{GMR_fig34} shows that for both dissipation models and both planets, the tidal dissipation reservoirs generally increase with the core radius (left) while they slightly decrease with increasing core mass --- or decreasing $\rho_o/\rho_c$ (right). These plots show that in Jupiter- and Saturn-like gaseous giant planets, the two distinct mechanisms exposed earlier can both contribute to tidal dissipation, and that therefore none of them can be neglected in general \citep[see][]{GMR2014}.
\end{itemize}

\section{Conclusions}

In the case of Jupiter and Saturn-like planets, we show that the viscoelastic dissipation in the core could dominate the turbulent friction acting on tidal inertial waves in the envelope. However, the fluid dissipation would not be negligible. This demonstrates that it is necessary to build complete models of tidal dissipation in planetary interiors from their deep interior to their surface without any arbitrary a-priori.

\bibliographystyle{aa}  
\bibliography{guenel2} 

\end{document}